\documentstyle[aps,prl,twocolumn]{revtex}
\begin{document}
\draft
\title{ Many-Spin Effects and Tunneling Properties of Magnetic Molecules }
\author{M.\ Al-Saqer, V.\ V.\ Dobrovitski, B.\ N.\ Harmon, and 
  M.\ I.\ Katsnelson}
\address{ Ames Laboratory, Iowa State University, Ames, Iowa 50011 }
\maketitle

\begin{abstract}
Spin tunneling in molecular magnets has attracted much attention, 
however theoretical considerations of this phenomenon
up to now have not taken into account the many-spin 
nature of molecular magnets. We present, to our knowledge, 
the first successful attempt of a realistic calculation of 
tunneling splittings for Mn$_{12}$ molecules, thus achieving
a quantitatively accurate many-spin description of a real 
molecular magnet in the energy interval ranging from about 
100 K down to 10$^{-12}$ K. Comparison with the results of the 
standard single-spin model shows that many-spin effects 
affect the tunneling splittings considerably. The
values of ground state splitting given by single-spin and many-spin
models differ from each other by a factor of five.
\end{abstract}


Progress in coordination chemistry lead recently to 
synthesis of a completely new class of magnetic materials of nanometer
size, molecular magnets, which has drawn the 
attention of physicists as well as chemists \cite{genintro}. 
In particular, these materials have been
proven to be very suitable for the study of mesoscopic 
quantum tunneling effects in magnetic materials. 
A number of impressive experimental results have been obtained 
recently, such as
the thermally-induced tunneling \cite{jumps}, 
ground state - to - ground state tunneling and
topological phase effects in spin tunneling \cite{fe8tun}.

Among others, the molecular magnet
$\rm Mn_{12}O_{12}(CH_3COO)_{16}(H_2O)_4$
(below referred to as Mn$_{12}$) constitutes a subject of great interest. 
The effect of resonant magnetization tunneling 
has been first observed and studied in detailed 
experiments on Mn$_{12}$ \cite{jumps}, and, at present, 
a substantial amount of reliable experimental data has been collected.
However, progress in understanding the properties of 
molecular magnets is greatly hampered by the lack 
of a comprehensive theoretical description. 
Recent experiments \cite{fe8tun,neutr}
show that the conventional single-spin description of magnetic 
molecules, including Mn$_{12}$, does not accurately describe 
their properties and the constituent many-spin nature of 
these molecules should be taken into account. This can be
particularly important for understanding tunneling properties;
for which it has been demonstrated \cite{spindyn} that the single-spin
description of a many-spin system can give totally misleading
results for tunneling splittings, differing by three orders 
of magnitude from exact values.

Therefore, the development of an adequate many-spin description 
of molecular magnets
constitutes an important theoretical problem, which becomes especially
complicated if the tunneling properties are to be studied.
For example, the magnetic cluster of Mn$_{12}$ molecule consists
of eight ions Mn$^{3+}$ with spin 2 and four Mn$^{4+}$ ions with 
spin 3/2 (see Fig.\ \ref{fig1}), coupled by 
anisotropic interactions, so that the Hilbert space
of the corresponding spin Hamiltonian consists of 10$^8$ levels.
Furthermore, the tunneling splittings in Mn$_{12}$ are very small,
of order of 10$^{-12}$ Kelvin. Obviously, the brute-force direct
calculation of tiny tunneling splittings, even for several low-lying
states, is far beyond the capabilities of modern computers.
Although several methods have been developed to cope with this problem
\cite{qmcmod,spindyn}, we are not aware of any calculations
of tunneling splittings made for realistic models of complex
molecular magnets, so that the possibility of such a simulation 
has not been demonstrated yet. 

In this paper, we present, to our knowledge, the first successful
attempt to calculate the tunneling splitting for a realistic
model of Mn$_{12}$ \cite{mn12ours}, thus explicitly showing
the possibility of calculating tunneling splittings 
in rather complex molecular magnets. As a result, we achieve
a quantitatively accurate many-spin description of a real system in the 
energy interval ranging from about 100 K down to 10$^{-12}$ K,
covering 14 orders of magnitude.

The 8-spin model, which forms a basis for present calculations,
has been described and discussed thoroughly in Ref.\ \onlinecite{mn12ours}.
The central idea of the model is to use the natural hierarchy of
interactions present in Mn$_{12}$. Namely, the antiferromagnetic
exchange interactions $J_1\simeq 220$ K (see Fig.\ \ref{fig1})
between ions Mn$^{3+}$ and Mn$^{4+}$ \cite{gat} are 
significantly stronger than other isotropic exchange ($J_2$, $J_3$
and $J_4$) and anisotropic interactions, so at low
temperatures they can be
considered as absolutely rigid. In other words, the spins of the
corresponding ions can be considered as forming dimers with a
total rigid spin 1/2. The
validity of this model has been supported by 
megagauss-field experiments \cite{dotsenko}: 
the states of dimers with the spin higher than 1/2 
(excitations of dimers) come 
into play when the external magnetic field 
is about 400 T. The same conclusion can be drawn from the 
the dependence of the magnetic susceptibility of Mn$_{12}$ 
versus temperature (see, e.g. Ref.\ \onlinecite{gat}).

After consideration of different possible interactions, the 
following spin Hamiltonian has been proposed in 
Ref.\ \onlinecite{mn12ours} to describe Mn$_{12}$ molecules:
\begin{eqnarray}
\label{hamilton}
{\cal H}&=& -J\Bigl(\sum_i {\bf s}_i \Bigr)^2
  -J'\sum_{\langle k,l\rangle} {\bf s}_k {\bf S}_l
  -K_z\sum_{i=1}^4 \left(S_i^z\right)^2\\ \nonumber
  && + \sum_{\langle i,j\rangle} {\bf D}^{i,j} \cdot
     [{\bf s}_i\times {\bf S}_j].
\end{eqnarray}
Here, ${\bf S}_i$ and ${\bf s}_i$ are the spin operators of
large spins $S=2$ and small dimer spins $s=1/2$, correspondingly,
where the subscript $i$ denotes the index of the spin.
Isotropic exchange between small spins and large spins is
described by the parameter $J'$, whereas $J$ denotes the
exchange of small spins with each other. The third term
describes the single-ion uniaxial anisotropy of large spins.
Finally, the fourth term describes the antisymmetric 
Dzyaloshinsky-Morya (DM) interactions in Mn$_{12}$,
and ${\bf D}^{i,j}$ is the Dzyaloshinsky-Morya vector describing
DM-interaction between $i$-th small spin and $j$-th large spin.
The molecules of Mn$_{12}$ possess the fourfold rotary-reflection
axis (the symmetry $C_4$), which imposes restrictions on
the DM-vectors ${\bf D}^{i,j}$,
so that Dzyaloshinsky-Morya interactions can be 
desribed by only three parameters $D_x \equiv D_x^{1,8}$,
$D_y\equiv D_y^{1,8}$, and $D_z \equiv D_z^{1,8}$.

The model discussed above has been demonstrated to describe correctly 
the energy spectrum up to
about 100 K \cite{mn12ours,trsusc}. It has allowed explaination of
a wide range of
experimental data, such as the unexpected splitting of neutron 
scattering peaks \cite{neutr}, results of EPR experiments 
\cite{hfepr} and measurements of magnetic susceptibility versus
temperature \cite{gat}. This model has provided a 
quantitative description of the response of Mn$_{12}$
molecules to the transversal magnetic field \cite{trsusc} 
(external field
applied perpendicular with respect to the easy axis of the
molecule). The following set of the parameters
has been determined for the spin Hamiltonian (\ref{hamilton}):
\begin{eqnarray}
&&J=3.6 \text{ K}, \quad J'=84 \text{ K}, \quad K_z=5.69 \text{ K}\\
  \nonumber
&&D_x=25.3 \text{ K}, \quad D_y=-0.6 \text{ K}, \quad D_z=-2.0 
  \text{ K}.
\end{eqnarray}
Having this model at hand, can we describe correctly
the tunneling properties of Mn$_{12}$? 

In the Hamiltonian (\ref{hamilton}), only the fourth term,
representing Dzyaloshinsky-Morya interactions, allows for
tunneling. Indeed, the first two terms (describing isotropic
exchange interactions) conserve both the total spin of the 
molecule $\cal S$ and its projection ${\cal S}_z$ (the
$z$-axis is chosen to coincide with the $C_4$ axis of the
molecule). Thus, these terms can not lead to tunneling
between the levels with different ${\cal S}_z$. The third 
term, representing an easy-axis anisotropy, also conserves
this quantity. Therefore, even though the levels $|{\cal S}_z=+M\rangle$
and $|{\cal S}_z=-M\rangle$ are degenerate, in the absense of the
Dzyaloshinsky-Morya term no tunneling between them
can appear. But the DM-interaction mixes levels with
different ${\cal S}_z$, thus giving rise to tunneling
of the molecule's spin. Below, for simplicity, we will
denote the energy levels by the value of ${\cal S}_z$. Although
it is not an exact quantum number in the
system under consideration, we can formally consider the
DM-interaction as a perturbation, and use 
perturbation theory terminology.

The first question to pose concerns the precision of the
level splitting calculation. Parameters of the Hamiltonian
are determined with some finite precision, and a small
error (say, of the order of several Kelvin) affects the
level energy by an amount of order of Kelvin, which is
much larger than the very small
value of tunneling splitting (of order of 10$^{-12}$ K).
Does it deprive the calculational results of all meaning?
To answer this question, we note that the levels 
$|{\cal S}_z=+M\rangle$ and $|{\cal S}_z=-M\rangle$ are
degenerate due to exact symmetry properties of the
spin Hamiltonian, and, in the absense of the DM-term,
would be degenerate at any values of parameters. Therefore,
the tunneling splittings $\Delta E_{+M, -M}$ are governed 
only by the strength
of the interaction which breaks the symmetry, i.e. DM-interaction.
Therefore, if the parameters of the Hamiltonian are determined
with reasonably small {\it relative\/} error, and if 
the simulation is done with sufficient accuracy, then the
{\it relative\/} error of the level splittings will 
also be small. The results of our calculations confirm this conclusion.

In the calculation scheme we employed \cite{mn12ours}, 
the most significant error
comes from the fact that only a fraction of all the levels produced 
by the spin Hamiltonian (\ref{hamilton}) are used in the calculations,
and the influence of levels having higher energy is neglected.
Therefore, to achieve reasonable accuracy, a sufficiently large number
of levels should be taken into account. We studied the dependence
of the resulting tunneling splittings $\Delta E_{+M, -M}$ 
for different pairs of degenerate levels $|{\cal S}_z=+M\rangle$ and 
$|{\cal S}_z=-M\rangle$ on the number of lowest levels
actually used in calculations. The results are shown on 
Fig.\ \ref{fig2}. It can be seen that the reasonable accuracy
can be achieved by accounting for about the 700 lowest levels.

Final results for the tunneling splittings $\Delta E_{+M, -M}$
are presented in the Table, where the values obtained in many-spin 
calculations
are compared with those given by the single-spin model. 
It is important to
mention that the tunneling splittings should be zero for the 
levels with odd values of $M$ (i.e., 
$|{\cal S}_z=\pm 9\rangle$, $|{\cal S}_z=\pm 7\rangle$ etc.)
because the fourfold symmetry of the molecule
imposes certain restrictions on the symmetry of the
spin Hamiltonian and makes some matrix elements vanish.
In the single-spin model of Mn$_{12}$ this property of the
spin Hamiltonian is introduced explicitly, by retaining only
those operators which possess the required fourfold symmetry.
In the many-spin simulations, we obtain the same result
independently:
the energies of the levels $|{\cal S}_z=\pm M\rangle$ for odd $M$ 
are the same with the accuracy of order of 10$^{-28}$ K, 
i.e. of order of computational
error. This value is much less than the smallest of 
the splittings $\Delta E_{+10,-10} = 2.03\cdot 10^{-12}$ K. 

Analyzing the results presented in the Table, we see that 
the single-spin and many-spin models give rather close
results for higher energy levels $M=\pm 2$ and $M=\pm 4$.
It has been shown \cite{single} that these values of 
tunneling splittings allow to describe the experimental
data available now with a good precision.
On the other
hand, for the ground-state levels $M=\pm 10$ the single-spin
model predicts the value of tunneling splitting five times
larger than the many-spin model. The possibility of such
a difference has been predicted before \cite{spindyn},
and, to our knowledge, Mn$_{12}$ is the first example of a
real system where this difference has been found.
Unfortunately, reliable experimental data concerning 
ground state - to - ground state tunneling in Mn$_{12}$
are absent, so the single-spin and many-spin models 
can not be distinguished on the basis of experimental results.
Nevertheless, many-spin consideration can be important
for a correct quantitative description of the 
ground state - to - ground state tunneling in other molecular magnets,
such as Fe$_8$. Many-spin effects may be one possible 
explanation for the disagreement found in
Ref.\ \onlinecite{fe8tun} between the experimental
results and predictions of the single-spin model.

Summarizing, we evaluated the tunneling splittings in
Mn$_{12}$ on the basis of the realistic many-spin model. 
We demonstrated that even tiny energy splittings, of order
of 10$^{-12}$ K, can be calculated with reasonable precision
for a rather complex molecular magnet. The results obtained
have been compared with predictions of the single-spin model.
We found that both models give close values for splittings of
higher levels. Thus, the models can not be distinguished on
the basis of experimental results available now, which provide
the information only about upper levels splittings \cite{single} 
($M=\pm 4$ and $M=\pm 2$). 
However, the ground state splittings calculated using
these two models differ by a factor of five. This difference
may be important also for other molecular magnets, such as Fe$_8$,
where the ground state splitting given by the single-spin model 
is three times less than the value obtained in experiment.

This work was carried out at the Ames Laboratory, which 
is operated for the U.\ S.\ Department of Energy by Iowa State 
University under Contract No.\ W-7405-82 and was supported by 
the Director for Energy Research, Office of Basic Energy Sciences 
of the U.\ S.\ Department of Energy.

\begin{figure}
\caption{Schematic plot of the Mn$_{12}$ cluster. Small black 
circles represent Mn$^{4+}$ ions, large white circles --- Mn$^{3+}$
ions. Different types of lines connecting the ions (solid, dashed,
dotted and dash-dotted) correspond to different types of isotropic
Heisenberg exchange interactions ($J_1$, $J_2$, $J_3$ and $J_4$).}
\label{fig1}
\end{figure}

\begin{figure}
\caption{Dependence of the logarithm of tunneling 
splittings $\Delta E_{+M, -M}$
versus the number of levels taken into account in the many-spin
calculations. The results for $M=10$, 8, 6, 4, and 2 are presented.
Tunneling splittings for the levels with odd $M$ are zero,
because of the symmetry properties of the spin Hamiltonian.}
\label{fig2}
\end{figure}

\begin{table}
\caption{Calculated values of tunneling splittings 
$\Delta E_{+M, -M}$: comparison
between the single-spin and the many-spin models.}
\begin{tabular}{ldd}
 $M$ & Many-spin model & Single-spin model \\
\tableline
$\pm 10$ & 2.03$\cdot 10^{-12}$ K & 1.25$\cdot 10^{-11}$ K \\
$\pm 8$ & 4.20$\cdot 10^{-8}$ K & 1.24$\cdot 10^{-7}$ K \\
$\pm 6$ & 9.18$\cdot 10^{-5}$ K & 1.20$\cdot 10^{-4}$ K \\
$\pm 4$ & 2.75$\cdot 10^{-2}$ K & 2.27$\cdot 10^{-2}$ K \\
$\pm 2$ & 8.51$\cdot 10^{-1}$ K & 7.01$\cdot 10^{-1}$ K \\
\end{tabular}
\label{tab}
\end{table}

\end{document}